\documentclass[acmtog]{acmart}
\acmSubmissionID{1537}

\usepackage{booktabs}
\usepackage[ruled]{algorithm2e}

\SetAlFnt{\small}
\SetAlCapFnt{\small}
\SetAlCapNameFnt{\small}
\SetAlCapHSkip{0pt}

\usepackage{multirow}
\usepackage{colortbl}
\usepackage{xcolor}
\definecolor{highlightrow}{RGB}{235, 245, 255}
\definecolor{headergray}{gray}{0.92}
\usepackage{tabularx}
\newcolumntype{Y}{>{\centering\arraybackslash}X}
\usepackage{arydshln}
\usepackage{enumitem}

\newcommand{\ie}{\emph{i.e.}}
\newcommand{\eg}{\emph{e.g.}}
\newcommand{\wrt}{\emph{w.r.t.}}

\acmJournal{TOG}
\begin{document}

\title{UniffordInteractor: Reasoning the Unified Affordance for Human and Multi-Object Interaction}

\author{Chen Su}
\affiliation{%
    \institution{Beijing University Of Posts and Telecommunications}
    \city{Beijing}
    \country{China}
}
\email{suchen1201@bupt.edu.cn}

\author{Xingqun Qi}
\affiliation{%
    \institution{Hong Kong University of Science and Technology}
    \city{HongKong}
    \country{HongKong}
}
\email{xingqunqi@gmail.com}

\author{Xurui Hu}
\affiliation{%
    \institution{Beijing University Of Posts and Telecommunications}
    \city{Beijing}
    \country{China}
}
\email{huxurui7@gmail.com}

\author{Bingkun Yang}
\affiliation{%
    \institution{Beijing University Of Posts and Telecommunications}
    \city{Beijing}
    \country{China}
}
\email{zhihenxue@bupt.edu.cn}

\author{Muyi Sun}
\authornote{Corresponding author.}
\affiliation{%
    \institution{Beijing University Of Posts and Telecommunications}
    \city{Beijing}
    \country{China}
}
\email{muyi.sun@bupt.edu.cn}

\author{Man Zhang}
\affiliation{%
    \institution{Beijing University Of Posts and Telecommunications}
    \city{Beijing}
    \country{China}
}
\email{zhangman@bupt.edu.cn}

\author{Zhenan Sun}
\affiliation{%
    \institution{Institute of Automation, Chinese Academy Of Sciences}
    \city{Beijing}
    \country{China}
}
\email{znsun@nlpr.ia.ac.cn}

\setcopyright{cc}
\setcctype{by}
\acmJournal{TOG}
\acmYear{2026} \acmVolume{45} \acmNumber{6} \acmArticle{193}
\acmMonth{12} \acmDOI{10.1145/3842530}

\begin{abstract}
Deriving plausible and dynamic Human Object Interaction (HOI) has gained tremendous progress in Embodied AI and Immersive AR/VR. 
While existing HOI methods have effectively explored human interaction with individual or fixed objects, they still show limitations in complex multi-object interaction scenes. 
Meanwhile, previous approaches focus more on the \textbf{contact information} held by human-object relation, overlooking the significance of the contact situation of various entities (\ie, object-object).
To address these limitations, we propose \textbf{UniffordInteractor}, a Reasoning-then-Interacting framework that leverages unified affordance for realistic and long-horizon human and multi-object interaction modeling. 
At the \textbf{reasoning} stage, we devise a \textbf{Unified Affordance Reasoner (UAR)} that employs a fine-tuned 3D Multimodal Large Language Model for unified object-centric affordance prediction. 
At the \textbf{interacting} stage, we propose \textbf{Hierarchical Interaction Generator (HIG)}, 
which formulates the interaction generation as a hierarchical (\eg, root$\rightarrow$body$\rightarrow$hand$\rightarrow$objects) autoregressive prediction, conditioned on the obtained unified affordance.
Extensive experiments demonstrate that UniffordInteractor surpasses SOTA counterparts in human and multi-object interaction generation.
Our project page is available at this \href{https://hermionee.github.io/UniffordInteractor/}{\textcolor{blue}{site}}. Code and dataset will be released upon publication.

\end{abstract}

\ccsdesc[500]{Computing methodologies~Animation}
\ccsdesc[300]{Computing methodologies~Shape modeling}
\ccsdesc[300]{Computing methodologies~Computer vision}

\keywords{Human-Object Interaction, multi-object interaction, affordance modeling, motion synthesis}

\begin{teaserfigure}
    \centering
    \includegraphics[width=\textwidth]{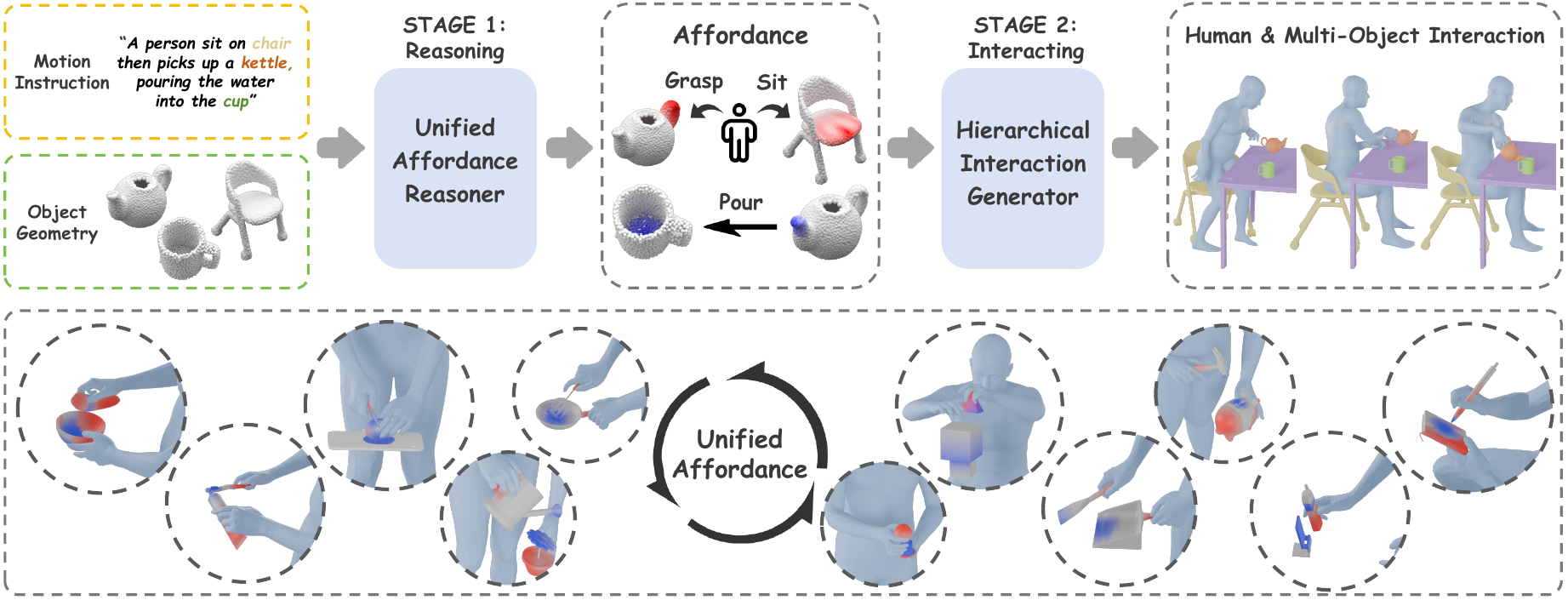}
    \caption{Overview of \textbf{UniffordInteractor}. 
We propose a \textbf{Reasoning-then-Interacting} framework for generating realistic human and multi-object interactions. 
Given a text instruction and object geometry, the Unified Affordance Reasoner first predicts \textbf{unified affordance} maps, 
explicitly modeling the interactive relations of both human-object (\textcolor{red}{red}) and object-object (\textcolor{blue}{blue}). 
Subsequently, the Hierarchical Interaction Generator synthesizes coherent 3D motions conditioned on these affordances.}
    \label{fig:teaser_image}
\end{teaserfigure}

\maketitle

\section{Introduction}
Human-Object Interaction (HOI) synthesis refers to generating plausible and dynamic movements of both humans and manipulable objects, 
which plays a crucial role of human behavior analysis deeply embedded in daily activities. 
HOI presents increasing attention in various human-centric applications such as Virtual/Augmented Reality (VR/AR)~\cite{guzov24ireplica}, Embodied AI (EAI), robotics~\cite{chen2025rhino,pan2022algorithms}, and avatar animation~\cite{cao2024avatargo,kim2016retargeting}. In recent years, numerous works have been conducted to address this challenging task.

Pioneering works in HOI usually focus on generating human interactions with individual objects, conditioned on object trajectories~\cite{li2023object, zhang2025diffgrasp} or textual descriptions~\cite{li2024controllable, zeng2025chainhoi, xue2025guiding}. Meanwhile, to enhance physical plausibility, a body of studies actively incorporates contact information to impose explicit constraints on interaction~\cite{song2024hoianimator, peng2025hoi, cong2025semgeomo}. Only few researchers attempt to explore multi-object scenarios by imposing geometric distance constraints~\cite{lv2024himo} or trajectory synchronization~\cite{he2025syncdiff}, which treat the entity relations as implicit kinematic byproducts primarily. However, these approaches ignore explicit contact constraints implied by interaction semantics.
This limitation is critical in multi-object scenarios, where objects exposes distinct affordance regions corresponding to different interacting partners. 
For instance, given the instruction ``\texttt{pick up a kettle and pour the water into the cup}": the motion requires maintaining physical contact with the kettle's handle (human-object) while simultaneously ensuring precise spatial alignment of the spout with the cup (object-object), as illustrated in Figure~\ref{fig:teaser_image}. 
Individual prediction of human-object or object-object affordances may generate local plausibility, but cannot ensure global alignment between multiple entities.
In addition, the spatial inconsistency draws an undesirable impact on temporally synchronized and physically aligned motion.

Thus, the unified and aligned interaction relations display indispensable roles in HOI generation. This highlights two main challenges: 1) The difficulty of uniformly modeling diverse interactions, as human-object and object-object relations involve different contact types. 2) The challenge of generating synchronized motions that keep all entities physically aligned over time.


To address these challenges, we propose UniffordInteractor, a Reasoning-then-Interacting framework that leverages unified affordance priors to enable realistic and long-horizon human and multi-object interaction generation.
Inspired by the recent success of 3D Multimodal Large Language Models (3D-MLLMs) in understanding and grounding spatial geometries~\cite{qi2024shapellm, yu2025seqafford, chu20253d}, at the Reasoning stage, we introduce the Unified Affordance Reasoner (UAR) to perform explicit affordance reasoning via a pre-trained 3D-MLLMs.
Different from previous works~\cite{li2024laso, chu20253d} that just focus on isolated affordance for a single object, 
we extend the affordance grounding task to human and multi-object interaction scenarios. 
Specifically, our framework could simultaneously predict the identity of the interaction target and a matched pair of affordance regions on the interacting objects.
To support this, we construct a dataset of over 13k interaction pairs. 
Crucially, this dataset unifies Human-Object and Object-Object interactions into a standard format, annotating the interaction target, functional roles, and their corresponding affordance.

To synthesize coherent long-horizon interactions, we propose a Hierarchical Interaction Generator (HIG) in the Interacting stage, building upon an autoregressive transformer framework. 
Here, to address the disparity between global dynamics and fine-grained manipulation, we employ hierarchical decoupled quantization to disentangle motion into root, body, and hand levels. 
This strategy effectively mitigates the interference between distinct motion scales, enabling the model to capture precise hand-object contacts without sacrificing whole-body coordination. 
Specifically, we introduce an Object-Centric Interaction Adapter (OCIA), which fuses object geometry with unified affordance priors into geometric tokens as a prefix condition for the autoregressive generator. 
Driven by this prefix and leveraging the decoupled representation, we implement a hierarchical autoregressive strategy that generates tokens in a sequence of root$\rightarrow$body$\rightarrow$hand$\rightarrow$objects. 
This structured generation flow ensures that the interaction is spatially anchored by the affordance priors, while intrinsically modeling the causal chain of interaction through a coarse-to-fine paradigm.


Overall, our contributions are summarized as follows:
\begin{itemize}[leftmargin=*]
    \item We propose UniffordInteractor, a Reasoning-then-Interacting framework that leverages unified affordance priors to enable realistic, \textcolor{black}{kinematically coherent}, and long-horizon human and multi-object interaction generation.
    \item We introduce the Unified Affordance Reasoner (UAR), which realizes a novel unified affordance reasoning strategy with the help of fine-tuned 3D-MLLMs.
    \item We design the Hierarchical Interaction Generator (HIG), which formulates the human and multi-object generation as a hierarchical autoregressive prediction, conditioned on the unified affordance. 
    \item Extensive experiments show that our method outperforms state-of-the-art counterparts, displaying high-fidelity and geometrically consistent interactive motions in complex multi-object scenarios.
\end{itemize}
\vspace{-0.6em}
\section{Related Work}

\begin{figure}
    \centering
    \includegraphics[width=0.92\linewidth]{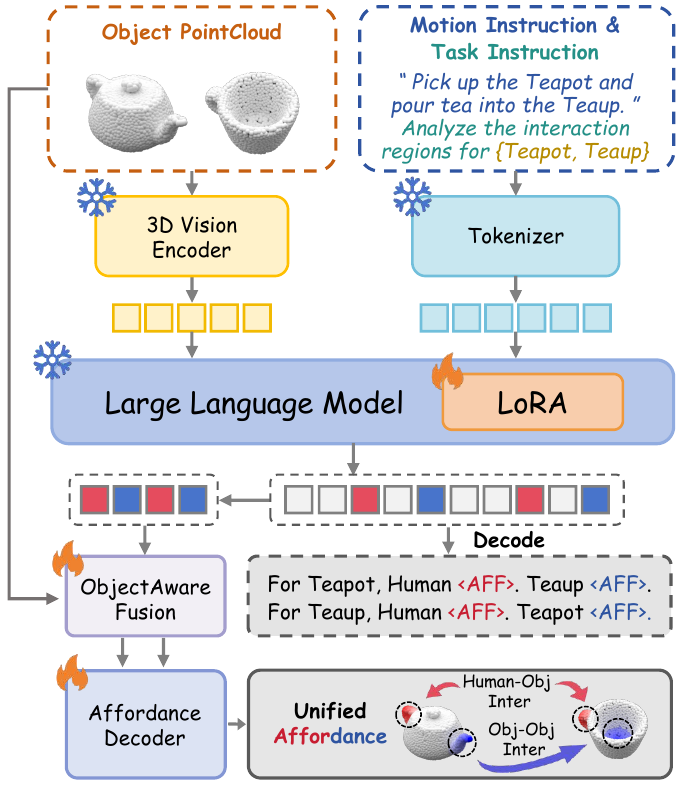}
    \vspace{-0.75em}
    \caption{Unified Affordance Reasoner (UAR). Given object point clouds and instructions, UAR reasons about interaction targets and predicts the corresponding affordance maps for each source object.}
    \label{fig:uar}
    \vspace{-0.5em}
\end{figure}

\textbf{Human-Object Interaction Generation:} Early works~\cite{starke2019neural,hassan2021stochastic,zhang2022couch} focus on generating interactions with static objects through a limited set of interaction modes, like sitting and lying. Others~\cite{li2023object, zhang2025diffgrasp} conditioned human motion on given object trajectories, failing to jointly synthesize both motions. 
InterDiff~\cite{xu2023interdiff} firstly proposes a diffusion-based framework to predict whole-body interactions with dynamic objects conditioned on a historical sequence. To generate more controllable interaction, a series of works~\cite{diller2024cg,li2024controllable,song2024hoianimator,peng2025hoi,cong2025semgeomo,zeng2025chainhoi,xue2025guiding} leverage textual descriptions to guide human–object interaction. Among them, HOI-Diff~\cite{peng2025hoi} enforces contact at predicted keypoints via a guidance mechanism. 
Recently, a few works have begun to explore the generation of human interactions with multiple objects. HIMO~\cite{lv2024himo} presents a benchmark for full-body human–multi-object interactions and a dual-branch conditional diffusion model for generating human and object motions. SyncDiff~\cite{he2025syncdiff} frames the task as multi-body motion synthesis, jointly modeling individual and relative motions with diffusion under synchronization constraints. However, these methods mainly model multi-entity relations at the kinematic level, omitting semantic contact constraints that result in ambiguous or physically inconsistent contacts.

\textbf{Affordance Learning:} Affordance learning initially focuses on 2D images~\cite{do2018affordancenet, chuang2018learning}, aiming to detect object-level affordance regions. Subsequent research~\cite{fang2018demo2vec,nagarajan2019grounded,luo2022learning} leaps to model affordances at the part-level, allowing more accurate identification of functional regions. Driven by embodied AI, research shift to 3D environments. 3D AffordanceNet~\cite{deng20213d} introduced the first point cloud benchmark based on PartNet~\cite{mo2019partnet}. Building on this, some researchers~\cite{yang2023grounding} attempt to transfer interaction priors from 2D HOI priors to 3D segmentation. Nevertheless, these approaches typically infer affordances purely from visual information, overlooking the semantic dimension. LASO~\cite{li2024laso} proposed language-conditioned segmentation to ground textual intent, while 3D AffordanceLLM~\cite{chu20253d} leveraged LLMs for open-vocabulary reasoning. Yet, these approaches are largely confined to isolated objects. While SeqAfford~\cite{yu2025seqafford} attempts to reason multi-object affordance, it independently predicts every object affordance in a sequential manner. Our framework jointly identifies the interaction counterpart and predicts a matched pair of affordance regions on the interacting objects.

\section{Method}

\subsection{Problem Formulation}
Given a motion instruction $\mathcal{I}$ and a set of 3D object geometries $\mathcal{G} = \{G_1, \dots, G_N\}$, where $N$ denotes the number of objects, our UniffordInteractor framework aims to synthesize \textcolor{black}{kinematically coherent} and temporally synchronized human motion $\mathcal{H}$ along with multiple object motions $\mathcal{O} = \{O_1, \dots, O_N\}$. The overall objective
is expressed as
\begin{equation}
    (\mathcal{H}, \mathcal{O}) = \text{UniffordInteractor}(\mathcal{I}, \mathcal{G}). 
\end{equation}



\textbf{Data Representation.} We leverage the SMPL-X~\cite{pavlakos2019expressive} parametric model to represent the human motion, where the human state in each frame is defined over 52 joints, consists of 3D joint positions, continuous 6D joint rotations, and global root translation. 
Meanwhile, for each object, the state is defined by global 6D rotation, 3D translation, and 2D object articulation parameters. And for rigid objects, the articulation term is omitted.
Specifically, we instantiate each object geometry $G_k \in \mathcal{G}$ as a sampled point cloud $P \in \mathbb{R}^{2048 \times 3}$, which serves as one input for both affordance reasoning and interaction generation.

\subsection{Unified Affordance Reasoning}
A critical limitation in human and multi-object interaction lies in the isolated treatment of human manipulation, which overlooks the mutual constraints between interacting objects. 
\textcolor{black}{To address this, we introduce Unified Affordance (Unifford), defined as the set of spatial interaction regions on a source object, corresponding to the interacting entities within a specific interaction task. 
For Human-Object interactions, the object is always designated as the source by definition, as the affordance characterizes a functional region on the object surface. For Object-Object interactions, the source is the object that initiates the interaction and the target is the interacting counterpart. For example, in contain, the apple initiates the containment interaction with the bowl, so the apple is the source and the bowl is the target.}
Crucially, this Unifford representation unifies diverse interaction types, which seamlessly integrates the human manipulation region and the aligned object-object geometric regions into a consistent spatial format.

\begin{figure*}[t]
    \centering
    \includegraphics[width=0.88\textwidth]{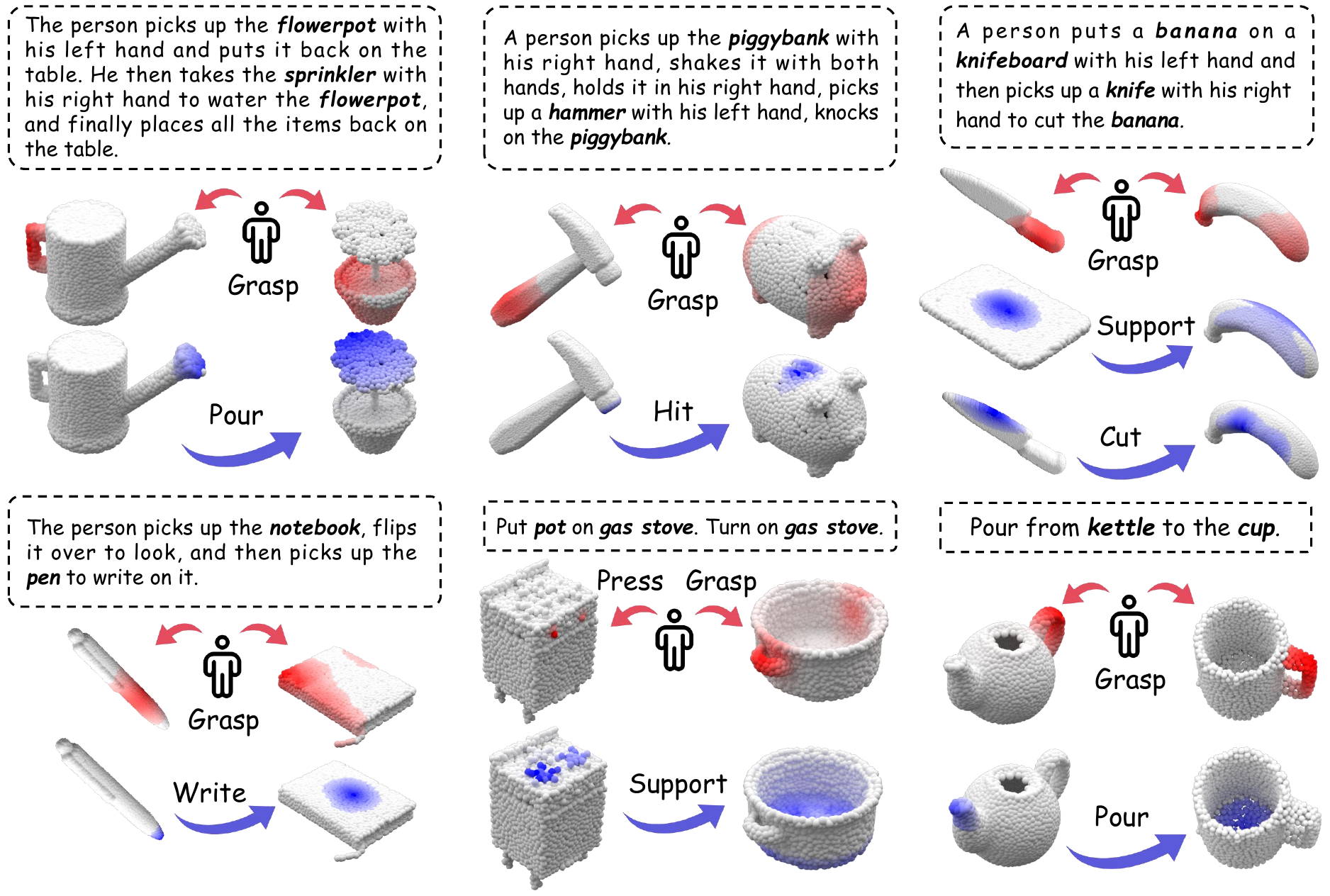}
    \vspace{-0.9em}
    \caption{Representative samples of InterAfford annotations. We visualize the textual descriptions and their corresponding dense affordance maps.}
    \label{fig:interafford2}
\end{figure*}

Specifically, to disentangle specific spatial contact constraints from complex task instructions, we propose the \textbf{Unified Affordance Reasoner (UAR)}, which leverages a fine-tuned 3D-MLLM to perform explicit affordance reasoning.
We introduce a newly designed token \texttt{\textless AFF\textgreater} into the original LLM vocabulary, that endow the affordance prediction capability into the LLM.
Formally, given a set of point clouds and a text instruction consisting of motion instruction and task instruction, the model is trained to predict an interleaved identity-affordance sequence $\mathcal{Y} = \langle I_1, \texttt{\textless AFF\textgreater}_1,
I_2, \texttt{\textless AFF\textgreater}_2, \ldots \rangle$, where each $\texttt{\textless AFF\textgreater}_k$ decodes the affordance region $A_k$ corresponding to the interaction target $I_k$. 
As shown in Figure~\ref{fig:uar}, we adopt ShapeLLM~\cite{qi2024shapellm} as our backbone, comprising a 3D vision encoder ReCon++~\cite{qi2024shapellm} to capture fine-grained object geometries and a LLaMA-based large language model to interpret these spatial features and reason complex interactions.
After the LLM generates the interleaved sequence, the \texttt{\textless AFF\textgreater} tokens are used as semantic queries. The Object-Aware Fusion module aligns these queries $q$ with the source object's geometry via cross-attention and calculates attention-based relevance scores between these queries and the geometric features of context objects $f_{ctx}$ to derive an aggregated representation $f_{agg}$:
\begin{equation}
    f_{agg}=\sum_{j=1}^M\mathrm{Softmax}(\mathrm{score}(q,f_{ctx}^j))f_{ctx}^j
\end{equation}
Guided by this aggregated representation, the Affordance Decoder predicts context-specific affordance regions by adaptively balancing intrinsic object properties with extrinsic environmental constraints.

\textbf{Two-stage Curriculum training:} To equip UAR with fine-grained grounding and relational reasoning, we initially fine-tune the model on a large-scale single-object affordance dataset~\cite{yu2025seqafford} for accurate mapping from language to spatial masks.
We then fine-tune it for Unifford reasoning in dynamic multi-object manipulation scenarios.
The framework is optimized with a composite objective, applying auto-regressive cross-entropy loss $\mathcal{L}_{\mathrm{ce}}$ to the generated interleaved sequence $\hat{\mathcal{Y}}$ and a combined Binary Cross-Entropy and Dice loss to the predicted affordance $\hat A$.
\begin{equation}
\begin{aligned}
\mathcal{L}_{\mathrm{uar}} = \lambda_{\mathrm{ce}} \mathcal{L}_{\mathrm{ce}}(\mathcal{Y}, \hat{\mathcal{Y}})
{}&+ \lambda_{\mathrm{bce}} \mathcal{L}_{\mathrm{bce}}(A, \hat{A}) \\
{}&+
\lambda_{\mathrm{dice}} \mathcal{L}_{\mathrm{dice}}(A, \hat{A}),
\end{aligned}
\end{equation}
where the weights $\lambda_{\mathrm{ce}},\lambda_{\mathrm{bce}},\lambda_{\mathrm{dice}}$ are utilized to balance the different loss items.


\subsection{InterAfford Dataset Construction}
\label{sec:inferafford}
To support the relational reasoning and dynamic multi-object manipulation scenarios mentioned above, we constructed an instruction-tuning dataset derived from high-fidelity interaction sequences~\cite{lv2024himo,kim2025parahome}, dubbed InterAfford. This dataset provides the necessary semantic-geometric grounding by unifying Human-Object and Object-Object interactions into a consistent format. The construction pipeline consists of the following key stages:

\textbf{Hybrid Affordance Annotation.} To capture diverse interaction types, we employ a hybrid strategy for generating dense 3D affordance maps. 
For contact-based interactions, we calculate the vertex-level distances between the human and the object. \textcolor{black}{A vertex is marked as contacting when the distance is below 2 cm. We accumulate the binary contact indicators over the interaction interval and normalize by the number of valid frames to obtain a per-vertex contact-frequency map.}
\textcolor{black}{For non-contact interactions such as pouring, we use the pre-trained LASO model~\cite{li2024laso} to generate pseudo-affordance maps. Annotators inspect the predictions with a point-cloud annotation tool, remove incorrect regions, and refine missing or inaccurate functional regions.}

\begin{figure}
    \centering
    \includegraphics[width=0.98\linewidth]{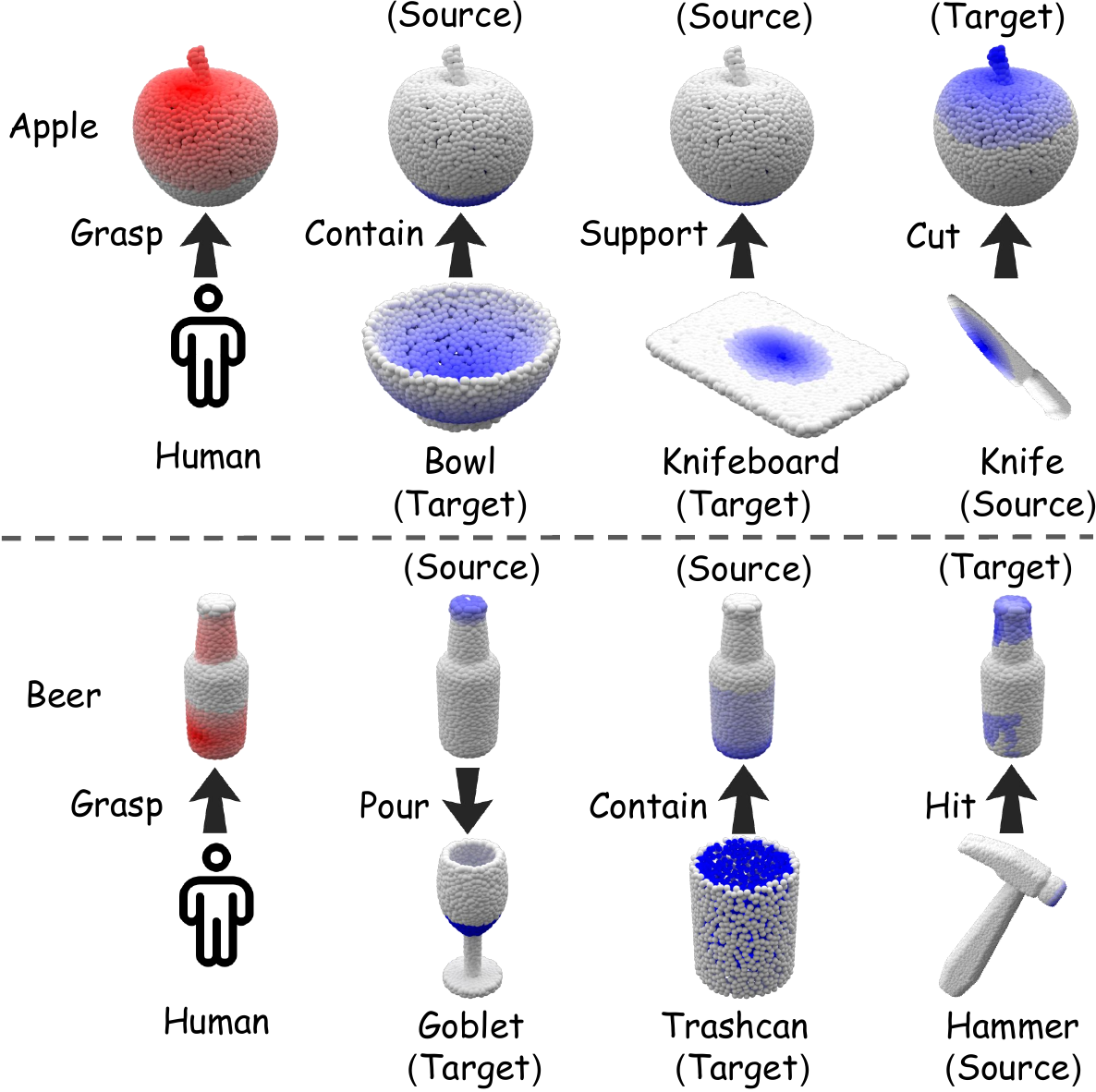}
    \vspace{-0.9em}
    \caption{Visualization of context-dependent affordances. A single object exhibits distinct roles and contact regions depending on the interacting entity.}
    \label{fig:interafford1}
\end{figure}

\textbf{Semantic Parsing and Augmentation.} For semantic processing, we leveraged a Large Language Model (LLM) to parse raw interaction descriptions, explicitly extracting the affordance labels and identifying interaction relationships between entities (source object and target). Following automated parsing, a manual verification process was conducted to ensure alignment between the visual affordance and textual semantics. To enhance linguistic diversity during instruction tuning, we employed LLM to augment
the interaction descriptions. Specifically, each original text description was rewritten into five distinct variations, ensuring the model
generalizes well across different instructional prompts.

\textbf{Dataset Statistics.} Through the construction pipeline described above, we obtained a total of 13,719 interaction pairs. Each data entry consists of key attributes including a \textbf{caption pool} of linguistically diverse instructions, the identities of the \textbf{interaction entities} (specifically the \textbf{source object} and \textbf{interaction target}), the \textbf{interaction type} (Human-Object or Object-Object), and the corresponding \textbf{affordance label}. 
To clarify the dynamic relationship, we explicitly mark the \textbf{role} and provide a dense 3D \textbf{affordance map} grounded to the object's geometry.
In Figure~\ref{fig:interafford2}, we present representative samples to demonstrate the explicit correspondence between the textual descriptions and the ground-truth affordance maps. Furthermore, to highlight the context-dependent versatility, we visualize specific examples in Figure~\ref{fig:interafford1}, where a single object (e.g., an Apple or a beer) exhibits distinct affordance maps and roles depending on the interacting entity.

\subsection{Hierarchical Interaction Generator}
\label{sec:hig}
Given the structured interaction cues from UAR, the subsequent goal is to synthesize \textcolor{black}{kinematically coherent} and temporally synchronized interactions. 
As shown in Fig.~\ref{fig:hig}, we introduce the Hierarchical Interaction Generator (HIG), which consists of two stages: Hierarchical Decoupled Quantization (Sec.~\ref{sec:hdq}) discretizes continuous motion into multi-scale tokens, and Hierarchical Autoregressive Modeling (Sec.~\ref{sec:ar}) synthesizes interactions via a coarse-to-fine autoregressive process.

\subsubsection{Hierarchical Decoupled Quantization}
\label{sec:hdq}
\begin{figure*}[t]
    \centering
    \includegraphics[width=0.96\textwidth]{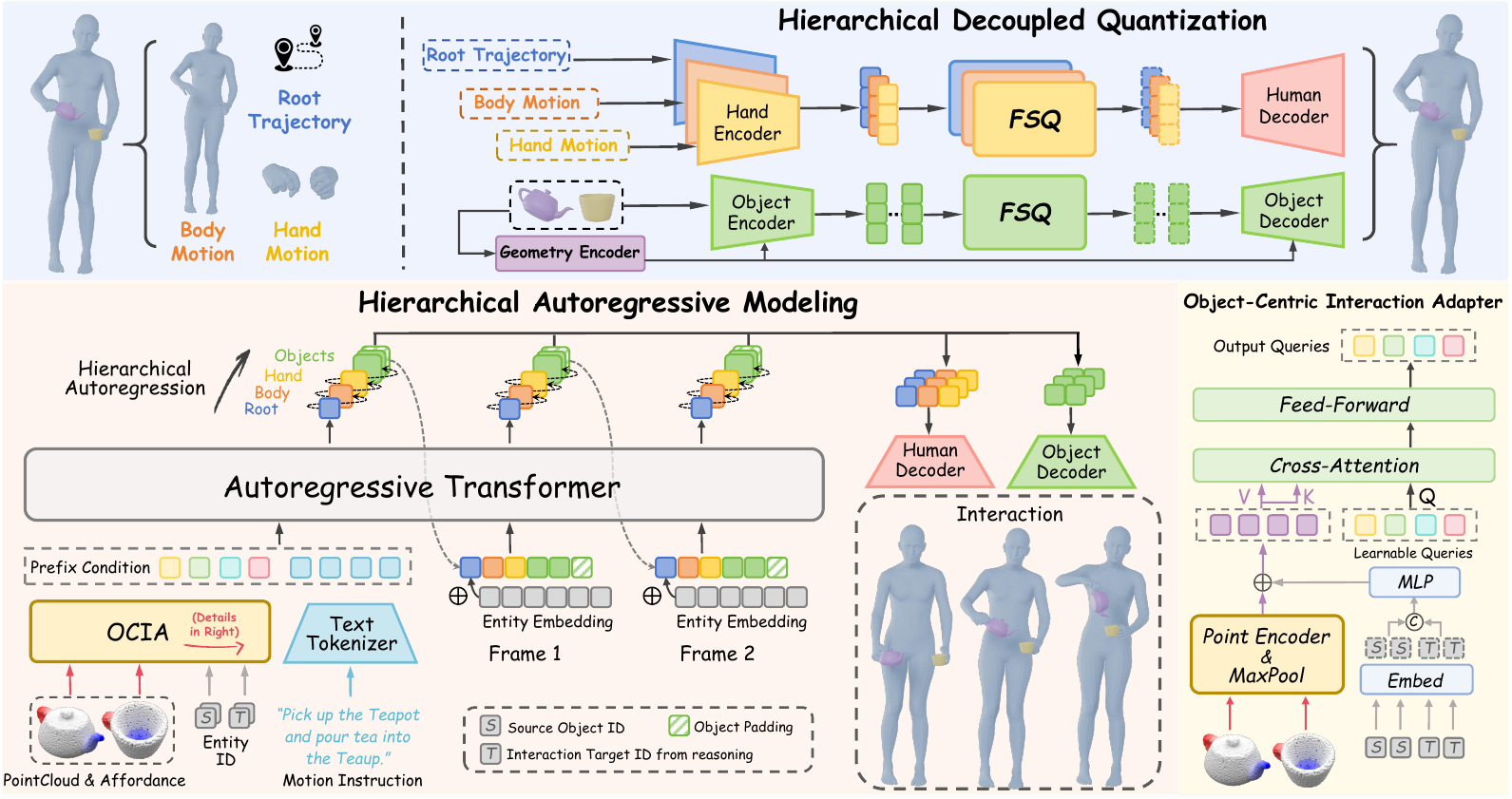}
    \vspace{-0.9em}
    \caption{The two-stage Hierarchical Interaction Generator for human and multi-object interaction. In stage one, we apply hierarchical decoupled quantization to discretize continuous human and object motions into multi-level discrete tokens. In stage two, we perform hierarchical autoregressive modeling in a coarse-to-fine manner to synthesize coherent human and multi-object interaction sequences.}
    \label{fig:hig}
    \vspace{-0.9em}
\end{figure*}

Modeling human-object interactions via a holistic representation is often suboptimal due to the fundamental structural disparities: human motion follows a complex skeletal topology, whereas objects exhibit diverse kinematic structures. To address this, we propose Hierarchical Decoupled Quantization, a dual-stream framework illustrated in Figure~\ref{fig:hig} (top).
For the human branch, to prevent large-scale global dynamics from overshadowing fine-grained hand details, we explicitly decompose motion into Root, Body, and Hand levels, encoding each into a dedicated codebook.
In parallel, the object branch adopts an instance-level encoding strategy that augments temporal object states with explicit geometric information. Specifically, a Geometry Encoder utilizing PointNet++~\cite{qi2017pointnet++} extracts geometry-aware keypoint representations from object point clouds. To enhance spatial expressiveness, we apply a high-frequency spatial encoding to the 3D keypoint coordinates, allowing the representation to capture fine-grained geometric variations before fusing it with rigid motion features.

Finally, we employ Finite Scalar Quantization (FSQ)~\cite{mentzer2023finite} to discretize these decoupled streams. Instead of nearest-neighbor lookup in an explicit codebook, FSQ constructs an implicit combinatorial codebook by bounding the continuous latent vector $z \in \mathbb{R}^d$ via a bounding function $f(\cdot)$ and rounding values to fixed integer levels:
\begin{equation}
    \hat{z} = \mathrm{round}(f(z)·(L-1))\in\{0,1,\dots,L-1\}^d
\end{equation}
where $L$ is the number of quantization levels per dimension. This per-dimension scalar quantization yields an exponentially large implicit codebook, effectively mitigating codebook collapse and ensuring high utilization for both hierarchical human features and instance-level object representations.


The model is optimized via a composite objective combining reconstruction loss $\mathcal{L}_{recon}$ and geometric interaction constraints $\mathcal{L}_{inter}$. Specifically, $\mathcal{L}_{inter}$ ensures \textcolor{black}{geometric consistency} through inter-object spatial alignment $\mathcal{L}_{sa}$, hand-object distance $\mathcal{L}_{dist}$, and penetration penalties $\mathcal{L}_{pen}$. Finally, the overall objective is formulated as:
\begin{equation}
\begin{aligned}[b]
    &\mathcal{L}_\mathrm{tokenize} = \mathcal{L}_\mathrm{recon}+\lambda_\mathrm{sa}\sum_{i<j} \|\Delta \hat{v}_{ij} - \Delta v_{ij}\|^2 \\
    \quad&+ \lambda_\mathrm{dist} \|\hat{\mathcal{D}} - \mathcal{D}\|^2 +\lambda_\mathrm{pen} \sum_{p \in P_{\mathrm{h}}} \left[ \max(0,\tau - d_p) \right]^2
\end{aligned}
\end{equation}
where $\Delta v_{ij}$ is the relative vertex distance, $\mathcal{D}$ is the hand-object distance matrix, $d_p$ denotes the shortest Euclidean distance from hand vertex to the object surface, and $\tau$ is the collision safety margin.

\subsubsection{Hierarchical Autoregressive Modeling}

\label{sec:ar}
Building on the discrete tokens produced by Hierarchical Decoupled Quantization, as illustrated in Figure~\ref{fig:hig} (bottom), Hierarchical Autoregressive Modeling captures probabilistic dependencies across multi-level components.
To efficiently handle a variable number of interacting objects within a unified batch, we design a unified frame layout that constructs a fixed-capacity token block for each timestep.
Specifically, each frame is formalized as a standardized sequence $[z_{t}^{root},z_{t}^{body},z_{t}^{hand},z_{t}^{obj_{1}},\ldots,z_{t}^{obj_{K}}]$, where $K$ denotes the maximum supported number of objects and unused slots are masked with \texttt{\textless PAD\textgreater} token. 
Crucially, since standard positional encodings lack semantic distinction for these diverse components, we inject a unique Entity ID into each slot, thereby enabling distinction between hierarchical body parts and independent object instances, explicitly.

\textbf{Object-Centric Interaction Adapter:} To convert the structured cues from UAR into actionable guidance, we propose the Object-Centric Interaction Adapter (OCIA) to extract and inject interaction-aware features. 
As illustrated in Figure~\ref{fig:hig} (lower right corner), OCIA encodes each object’s point cloud together with its affordance to capture geometry and contact-relevant regions.
\textcolor{black}{Here, source and target are structural labels denoting the two ends of an ordered interaction pair, used solely for relation encoding and independent of semantic role assignments.}
For relation modeling, the source object and its interaction targets are first mapped to predefined Entity IDs. To explicitly capture directional interaction semantics, we encode each ordered source–target pair by fusing the source embedding $\mathbf{e}_{src}^{(i)}$ and target embedding $\mathbf{e}_{tgt}^{(i,j)}$ via a relational MLP:
\begin{equation}
    \mathbf{r}_{i,j}=\mathrm{MLP}(\text{Concat}(\mathbf{e}_{src}^{(i)}, \mathbf{e}_{tgt}^{(i,j)})).
\end{equation}
Subsequently, these interaction embeddings are combined with the geometric feature to form a structured slot sequence. This sequence is aggregated by a set of learnable queries through cross-attention mechanism, producing interaction-aware conditioning tokens. \textcolor{black}{These tokens are derived from UAR predictions at both trainging and inference stage.}

With the unified frame layout and OCIA, we formulate motion generation as a hierarchical autoregressive process conditioned on text and object-centric interaction tokens. 
Unlike holistic approaches that predict all motion components simultaneously, our hierarchical autoregressive strategy explicitly models the causal dependency across multi-level components. 
Consistent with the frame layout, we factorize the joint distribution $p(\mathbf{Z}_t \mid \mathcal{C})$ into a causal chain, where each component of $\mathbf{Z}_t$ is conditioned on $\mathcal{C}$ and its preceding parent levels.
By generating interaction in a coarse-to-fine order (Root$\to$Body$\to$Hand$\to$Objects ), the model enforces that global body movements drives the subsequent fine-grained hand-object interactions, leading to more stable and \textcolor{black}{kinematically coherent} interactions.
The model is trained by minimizing the negative log-likelihood of the ground-truth motion tokens under the above factorization:
\begin{equation}
    \mathcal{L}_\mathrm{ar} = - \sum_t \log p(\mathbf{Z}_t \mid \mathcal{C}).
\end{equation}

\begin{table*}[t]
\centering
\footnotesize
\renewcommand{\arraystretch}{1.1}
\caption{Quantitative comparisons with SOTA methods on \textbf{HIMO} and \textbf{Parahome} datasets. 
Gray bars indicate different dataset settings. 
$\uparrow$ denotes higher is better, $\downarrow$ denotes lower is better. 
\textbf{Bold} indicates the best result.}
\label{tab:comparison}

\begin{tabularx}{0.97\textwidth}{l *{8}{Y}}
\toprule

\multirow{2.5}{*}{Methods} & 
\multirow{2.5}{*}{FID$\downarrow$} & 
\multicolumn{3}{c}{R-Precision$\uparrow$} & 
\multirow{2.5}{*}{Diversity$\rightarrow$} & 
\multirow{2.5}{*}{MM-Dist$\downarrow$} & 
\multirow{2.5}{*}{C\%$\uparrow$} & 
\multirow{2.5}{*}{Coll\%$\downarrow$} \\

\cmidrule(lr){3-5} 
& & Top 1 & Top 2 & Top 3 & & & & \\ 
\midrule

\rowcolor{headergray} 
\multicolumn{9}{c}{\textbf{\textit{On the HIMO Dataset}}} \\ 
\cdashline{1-9} 
\addlinespace[3pt]

Real &
$0.001$ & $0.501$ & $0.728$ & $0.841$ & $12.759$ & $3.749$ & $0.503$ & $0.171$ \\

\midrule

CHOIS~\cite{li2024controllable} \textcolor{gray}{\scriptsize $_{ECCV'24}$} & 
$3.626$ & $0.256$ & $0.415$ & $0.522$ & $10.685$ & $6.757$ & $0.346$ & $0.223$ \\

HIMO-Gen~\cite{lv2024himo} \textcolor{gray}{\scriptsize $_{ECCV'24}$} & 
$7.642$ & $0.244$ & $0.398$ & $0.511$ & $12.036$ & $6.681$ & $0.284
$ & $0.184$ \\

HOI-Diff~\cite{peng2025hoi} \textcolor{gray}{\scriptsize $_{CVPR'25}$} &
$12.746$ & $0.178$ & $0.287$ & $0.402$ & $10.744$ & $7.993$ & $0.238$ & $0.203$ \\

ROG~\cite{xue2025guiding} \textcolor{gray}{\scriptsize $_{CVPR'25}$} & 
$10.321$ & $0.202$ & $0.368$ & $0.481$ & $10.807$ & $7.748$ & $0.289$ & $0.192$ \\

\textcolor{black}{HOIDiNi~\cite{ron2026hoidini}} \textcolor{gray}{\scriptsize $_{CVPR'26}$} &
\textcolor{black}{$1.790$} & \textcolor{black}{$0.332$} & \textcolor{black}{$0.532$} & \textcolor{black}{$0.650$} & \textcolor{black}{$12.176$} & \textcolor{black}{$4.303$} & \textcolor{black}{$0.306$} & \textcolor{black}{$0.192$} \\

\rowcolor{highlightrow}
\textbf{UniffordInteractor (Ours)} & 
$\mathbf{1.199}$ & $\mathbf{0.493}$ & $\mathbf{0.680}$ & $\mathbf{0.789}$ & $\mathbf{12.604}$ & $\mathbf{4.015}$ & $\mathbf{0.348
}$ & $\mathbf{0.181}$ \\

\midrule
\rowcolor{headergray} 
\multicolumn{9}{c}{\textbf{\textit{On the Parahome Dataset}}} \\ 
\cdashline{1-9} 
\addlinespace[3pt]

Real &
$0.001$ & $0.540$ & $0.769$ & $0.875$ & $11.249$ & $2.055$ & $0.319$ & $0.213$ \\

\midrule

CHOIS~\cite{li2024controllable} \textcolor{gray}{\scriptsize $_{ECCV'24}$} & 
$1.123$ & $0.523$ & $0.746$ & $0.840$ & $10.035$ & $2.847$ & $0.163$ & $0.074$ \\

HIMO-Gen~\cite{lv2024himo} \textcolor{gray}{\scriptsize $_{ECCV'24}$} & 
$0.954$ & $0.516$ & $0.745$ & $0.851$ & $10.914$ & $2.835$ & $0.164$ & $0.093$ \\

HOI-Diff~\cite{peng2025hoi} \textcolor{gray}{\scriptsize $_{CVPR'25}$} &
$3.068$ & $0.373$ & $0.584$ & $0.714$ & $10.737$ & $3.947$ & $0.117$ & $\mathbf{0.036}$\\

ROG~\cite{xue2025guiding} \textcolor{gray}{\scriptsize $_{CVPR'25}$} & 
$1.614$ & $0.460$ & $0.676$ & $0.788$ & $10.822$ & $3.246$ & $0.178$ & $0.071$ \\

\textcolor{black}{HOIDiNi~\cite{ron2026hoidini}} \textcolor{gray}{\scriptsize $_{CVPR'26}$} &
\textcolor{black}{$0.612$} & \textcolor{black}{$0.424$} & \textcolor{black}{$0.680$} & \textcolor{black}{$0.797$} & \textcolor{black}{$10.721$} & \textcolor{black}{$3.115$} & \textcolor{black}{$\mathbf{0.205}$} & \textcolor{black}{$0.077$} \\

\rowcolor{highlightrow}
\textbf{UniffordInteractor (Ours)} & 
$\mathbf{0.536}$ & $\mathbf{0.520}$ & $\mathbf{0.755}$ & $\mathbf{0.866}$ & $\mathbf{10.990}$ & $\mathbf{2.161}$ & $0.192$ & $0.049$ \\

\bottomrule
\end{tabularx}
\end{table*}

\section{Experiments}
\subsection{Dataset and Experimental Setting}

\begin{table*}[h]
\centering
\footnotesize
\renewcommand{\arraystretch}{1.2}
\caption{Quantitative comparison between our \textbf{unified} model and HIMO-Gen's \textbf{specialized} models. 
We strictly follow the evaluation protocols on different object settings. 
$\uparrow$ denotes higher is better, $\downarrow$ denotes lower is better, $\rightarrow$ denotes closer to real is better. 
\textbf{Bold} indicates the best result.}
\label{tab:himo_specialized}

\begin{tabularx}{0.97\textwidth}{l *{8}{Y}}
\toprule

\multirow{2.5}{*}{Methods} & 
\multirow{2.5}{*}{FID$\downarrow$} & 
\multicolumn{3}{c}{R-Precision$\uparrow$} & 
\multirow{2.5}{*}{Diversity$\rightarrow$} & 
\multirow{2.5}{*}{MM-Dist$\downarrow$} & 
\multirow{2.5}{*}{C\%$\uparrow$} & 
\multirow{2.5}{*}{Coll\%$\downarrow$} \\

\cmidrule(lr){3-5} 
& & Top 1 & Top 2 & Top 3 & & & & \\ 
\midrule

\rowcolor{headergray} 
\multicolumn{9}{c}{\textbf{\textit{On the 2-Object Setting}}} \\ 
\cdashline{1-9} 
\addlinespace[3pt]
Real &
$0.001$ & $0.461$ & $0.678$ & $0.783$ & $11.369$ & $3.609$ & $0.547$ & $0.145$ \\
\midrule

HIMO-2o & 
$1.340$ & $0.318$ & $0.531$ & $0.648$ & $\mathbf{11.208}$ & $4.349$ & $0.292$ & $0.173$ \\

\rowcolor{highlightrow}
\textbf{UniffordInteractor (Ours)} & 
$\mathbf{0.969}$ & $\mathbf{0.424}$ & $\mathbf{0.663}$ & $\mathbf{0.769}$ & $11.903$ & $\mathbf{3.904}$ & $\mathbf{0.351}$ & $\mathbf{0.161}$ \\

\midrule
\rowcolor{headergray} 
\multicolumn{9}{c}{\textbf{\textit{On the 3-Object Setting}}} \\ 
\cdashline{1-9} 
\addlinespace[3pt]
Real &
$0.001$ & $0.375$ & $0.550$ & $0.663$ & $9.524$ & $3.791$ & $0.408$ & $0.194$ \\
\midrule

HIMO-3o & 
$4.706$ & $0.194$ & $0.356$ & $0.463$ & $9.320$ & $4.843$ & $0.155$ & $\mathbf{0.211}$ \\

\rowcolor{highlightrow}
\textbf{UniffordInteractor (Ours)} & 
$\mathbf{2.422}$ & $\mathbf{0.344}$ & $\mathbf{0.506}$ & $\mathbf{0.613}$ & $\mathbf{9.527}$ & $\mathbf{4.061}$ & $\mathbf{0.307}$ & $0.227$ \\

\bottomrule
\end{tabularx}
\end{table*}

\textbf{Dataset Processing:}
In this paper, we leverage two human-object interaction datasets (\ie, HIMO~\cite{lv2024himo} and ParaHome~\cite{kim2025parahome}) to verify the effectiveness of our method. Regarding the HIMO dataset that incorporates both two-object and three-object interaction scenarios, we follow the official train/validation/test split and jointly use both two-object and three-object data for training.

Besides, ParaHome dataset captures human activities in a realistic household environment, which features sequential multi-object manipulations covering both rigid and articulated objects. We segment these long-term, multi-stage activities into atomic interaction clips based on temporal annotations. Subsequently, for each interaction clip, we retain the relevant objects involved in the interaction and standardize the scene by transforming all spatial coordinates to be relative to the human's root pose at the first frame. In total, the processed dataset consists of 5,476 clips where the training set comprises 31 subjects and the test set includes the remaining 7 subjects.

\textbf{Implementation Details:} \textcolor{black}{We use K=3 object slots in all experiments based on our dataset's scenarios. Longer scenes involving more than three objects are decomposed into temporally ordered subsequences, which are generated autoregressively.} In the reasoning stage, we adopt ShapeLLM-7B as the 3D MLLM backbone, freezing the pre-trained vision encoder. Here, LoRA~\cite{hu2022lora} is employed for efficient fine-tuning with the rank set to 8 and the alpha set to 16. For the hierarchical quantization, we adopt FSQ and crop all motion sequences to $N=64$ frames during training. The model encodes the motion into four distinct codebooks corresponding to the root, body, hand, and objects, with hidden dimensions configured as 128, 512, 512, and 128, respectively. The codebook sizes were set to $N_{root}=512$ and $N_{body/hand/objects}=4096$. The initial learning rate is set to 1e-4 with a batch size of 128. We set \{$\lambda_{sa}$, $\lambda_{dist}$, and $\lambda_{pen}$\} = \{1.0, 1.0, 0.5\}. The autoregressive transformer is built upon LLaMA-based architecture with approximately 775M parameters. We utilize a frozen T5-XL model as the text encoder to process the textual descriptions. 
To mitigate exposure bias, we employ token-level scheduled sampling, where the probability $\epsilon$ of substituting ground-truth tokens with model predictions is annealed from 0.0 (teacher forcing) to 1.0 (pure autoregressive) via a cosine schedule. \textcolor{black}{During inference, our full pipeline, including unified affordance reasoning and hierarchical interaction generation, takes approximately 8.27 seconds per sample.} All the experiments are conducted on 4 NVIDIA A100 GPUs.

\textbf{Evaluation Metrics.} To comprehensively assess the capabilities of our model, we conduct quantitative evaluations from three distinct perspectives: Motion Quality, Interaction Quality, and Diversity: 
\begin{itemize}[leftmargin=*]
    \item \textbf{Motion Quality:} Fr\'echet Inception Distance (\textbf{FID}) measures the distribution gap between real and generated motions. \textbf{R-Precision} and \textbf{MM-Dist} evaluate text-motion alignment, while \textbf{MPJPE} assesses reconstruction fidelity in the quantization stage.
    \item \textbf{Interaction Quality:} \textbf{C\%} means the proportion of frames with detected human–object contact, and \textbf{Coll\%} reports the proportion of frames with human–object collisions.
    \item \textbf{Diversity:} \textbf{Diversity} measures motion variability based on the variance of extracted motion features
\end{itemize}

\subsection{Quantitative Evaluation}
\textbf{Comparisons with SOTA Methods:} To evaluate the effectiveness of our framework, we compare it against \textcolor{black}{five} State-Of-The-Art (SOTA) methods on both the HIMO and ParaHome datasets. 
For single-object interaction methods, including CHOIS~\cite{li2024controllable}, HOI-Diff~\cite{peng2025hoi}, and ROG~\cite{xue2025guiding}, we re-implement them to support multi-object inputs. 
Since HIMO-Gen~\cite{lv2024himo} is originally designed with separate weights for varying object counts, we adapt it into a single unified network to align with our experimental setting. 

The results in Table~\ref{tab:comparison} show our UniffordInteractor outperforms all the SOTA counterparts in both HIMO and ParaHome benchmarks. 
We achieve the best FID and the highest R-Precision across all settings, highlighting its superiority in motion quality and instruction following. This indicates that our HIG stage effectively captures high-fidelity motion dynamics, while the UAR stage ensures that the generated behaviors are semantically aligned with complex interaction instructions in multi-object scenarios.
In terms of interaction quality, our method achieves a superior trade-off, maximizing contact precision while rigorously suppressing invalid penetrations. 
This stems from the Unifford, which geometrically constrains interactions to valid regions.

\begin{table}[t]
    \centering
    \color{black}
    \small 
    \renewcommand{\arraystretch}{1.2}
    \caption{\textcolor{black}{Quantitative comparison of object–object interaction quality on HIMO. OO-C\% and OO-Coll\% adopt the same protocols as C\% and Coll\%, respectively, computed over object-object pairs.}}
    \label{tab:oo_interaction}
    \begin{tabularx}{\columnwidth}{Xcc}
        \toprule
        \textbf{Methods} & \textbf{OO-C\%} $\uparrow$ & \textbf{OO-Coll\%} $\downarrow$ \\
        \midrule
        CHOIS & 0.224 & 0.135 \\
        HIMO-Gen & 0.179 & 0.126 \\
        HOI-Diff & 0.146 & \textbf{0.108} \\
        ROG & 0.161 & 0.145 \\
        HOIDiNi & 0.235 & 0.156 \\
        \rowcolor{highlightrow} 
        \textbf{Ours} & \textbf{0.238} & 0.133 \\
        \bottomrule
    \end{tabularx}
\end{table}

\begin{table}[t]
    \centering
    \color{black}
    \small 
    \renewcommand{\arraystretch}{1.2}
    \caption{\textcolor{black}{Quantitative comparison with existing affordance prediction methods on the InterAfford dataset. We report AUC, SIM, and IOU to evaluate the quality of predicted affordance maps.}}
    \label{tab:comparison_affordance}
    \begin{tabularx}{\columnwidth}{Xccc}
        \toprule
        \textbf{Methods} & \textbf{AUC} $\uparrow$ & \textbf{SIM} $\uparrow$ & \textbf{IOU} $\uparrow$ \\
        \midrule
        SeqAffod & 0.8040 & 0.4847 & 0.1861 \\
        3D-AffordanceLLM & 0.8008 & 0.5053 & \textbf{0.2078} \\
        \rowcolor{highlightrow} 
        \textbf{UAR (Ours)} & \textbf{0.8223} & \textbf{0.5170} & 0.2036 \\
        \bottomrule
    \end{tabularx}
\end{table}

\begin{table}[t]
    \centering
    \color{black}
    \small
    \renewcommand{\arraystretch}{1.2}
    \setlength{\tabcolsep}{4.5pt} 
    \caption{\textcolor{black}{OOD generalization evaluation on HIMO. The model is trained on 42 objects and evaluated on both seen (in-distribution) and 11 unseen (OOD) object geometries.}}
    \label{tab:ood}
    \begin{tabularx}{\columnwidth}{Xccccc}
        \toprule
        \textbf{Settings} & \textbf{FID} $\downarrow$ & \textbf{R-Pre@3} $\uparrow$ & \textbf{MM-Dist} $\downarrow$ & \textbf{C\%} $\uparrow$ & \textbf{Coll\%} $\downarrow$ \\
        \midrule
        In-distribution & 1.316 & 0.760 & 3.712 & 0.346 & 0.177 \\
        OOD             & 4.587 & 0.541 & 4.067 & 0.314 & 0.231 \\
        \bottomrule
    \end{tabularx}
\end{table}

\textbf{Comparison against Task-Specific Baselines:} Furthermore, to ensure a rigorous evaluation against task-specific distributions, we compare our unified framework with HIMO-Gen’s specialized variants: HIMO-2O and HIMO-3O, which are trained exclusively on 2-object and 3-object data, respectively. As shown in Table~\ref{tab:himo_specialized}, our method consistently outperforms both variants, despite their advantage of task-specific training. Notably, in the complex 3-object scenario, our approach achieves a near 50\% reduction in FID (2.422 vs. 4.706). This substantial margin demonstrates that our ``Reasoning-then-Interacting'' paradigm, driven by explicit object-object affordance modeling, is essential for managing the complexity of multi-entity interactions.

\textcolor{black}{\textbf{Comparison of Object-Object Interaction:} To directly evaluate the interaction quality between multiple objects, we further report OO-C\% and OO-Coll\% on the HIMO dataset. As shown in Table~\ref{tab:oo_interaction}, our method achieves the highest object–object contact percentage while maintaining a low collision percentage. Although several baselines obtain lower collision percentages, they also exhibit substantially lower object–object contact rates suggest that they may fail to realize the intended object-object interactions. These results demonstrate that the explicit object–object affordance in Unifford provides effective geometric guidance for aligning multiple interacting entities.}

\textcolor{black}{\textbf{Comparison with Affordance-Prediction Methods:} We further compare the proposed Unified Affordance Reasoner with existing affordance prediction methods, including SeqAfford and 3D-AffordanceLLM. As shown in Table~\ref{tab:comparison_affordance}, UAR achieves the best AUC and SIM, demonstrating more accurate and distributionally consistent affordance grounding. Although 3D-AffordanceLLM obtains a marginally higher IOU, it predicts affordance independently for each object without interaction context. In contrast, UAR explicitly incorporates all interacting objects as contextual cues, which is designed to promote consistent predictions across interaction participants.}

\begin{table*}[t]
\centering
\small
\renewcommand{\arraystretch}{1.25}
\caption{Ablation study on Hierarchical Autoregressive Modeling. ``w/o hierarchical autoregressive'' indicates using standard next-token prediction on a single sequence. ``w/o entity embedding'' indicates we remove entity embeddings added to each token. ``w/o affordance'' indicates removing object-centric interaction adapter. ``w/o Unifford'' indicates removing object-object affordance. \textcolor{black}{``cascaded generation'' first generates the root and body sequence, then generates the hand sequence conditioned on the fixed body motion, and finally generates object motions conditioned on the generated human motion.}}
\label{tab:ablation_autoregressive}

\begin{tabularx}{0.97\textwidth}{l Y YYY c Y Y Y}
\toprule
\multirow{2.5}{*}{Methods} & 
\multirow{2.5}{*}{FID$\downarrow$} & 
\multicolumn{3}{c}{R-Precision$\uparrow$} & 
\multirow{2.5}{*}{Diversity$\rightarrow$} & 
\multirow{2.5}{*}{MM-Dist$\downarrow$} & 
\multirow{2.5}{*}{C\%$\uparrow$} & 
\multirow{2.5}{*}{Coll\%$\downarrow$} \\
\cmidrule(lr){3-5} 
& & Top 1 & Top 2 & Top 3 & & & & \\ 
\midrule

w/o hierarchical autoregressive & 
11.860 & 0.141 & 0.249 & 0.329 & 8.455 & 8.273 & 0.066 & 0.195 \\

\textcolor{black}{cascaded generation} &
\textcolor{black}{3.380} & \textcolor{black}{0.402} & \textcolor{black}{0.620} & \textcolor{black}{0.726} & \textcolor{black}{12.131} & \textcolor{black}{4.752} & \textcolor{black}{0.256} & \textcolor{black}{0.188} \\

w/o entity embedding & 
4.865 & 0.369 & 0.552 & 0.640 & 12.396 & 5.276 & 0.192 & 0.185 \\

w/o affordance & 
2.190 & 0.312 & 0.497 & 0.613 & 12.271 & 5.404 & 0.242 & \textbf{0.179} \\

w/o Unifford &
1.618 & 0.385 & 0.574 & 0.682 & 12.421 & 4.732 & \textbf{0.366} & 0.201 \\

\rowcolor{highlightrow}
\textbf{Ours (full version)} & 
\textbf{1.199} & \textbf{0.493} & \textbf{0.680} & \textbf{0.789} & \textbf{12.604} & \textbf{4.015} & 0.348 & 0.181 \\

\bottomrule
\end{tabularx}
\end{table*}

\textcolor{black}{\textbf{Generalization to Unseen Objects:} To assess the generalization capability of our method to unseen object geometries, we conduct an out-of-distribution evaluation by training on 42 objects from the HIMO dataset and testing on 11 held-out objects unseen during training. As reported in Table~\ref{tab:ood}, our model exhibits expected degradation in motion quality (\eg, FID) under the distributional shift introduced by novel geometries, while preserving contact accuracy (C\%: 0.346→0.314) and text-motion alignment (MM-Dist: 3.712→4.067). This suggests that unified affordance reasoning retains partial generalization to unseen geometries, particularly in contact accuracy.}

\textbf{Ablation Study:} To verify the effectiveness, we conduct more ablation studies on the comprehensive HIMO dataset.

\begin{table}[t]
    \centering
    \small
    \renewcommand{\arraystretch}{1.1}
    \caption{Ablation study on different fine-tuning strategies for the Unified Affordance Reasoner (UAR). We report AUC, SIM, and IOU to evaluate the quality of predicted affordance maps.}
    \label{tab:ablation_mllm}
    \begin{tabular}{lccc}
        \toprule
        \textbf{Methods} & \textbf{AUC} $\uparrow$ & \textbf{SIM} $\uparrow$ & \textbf{IOU} $\uparrow$ \\
        \midrule
        w/o Fine-tuning & 0.4813 & 0.3699 & 0.1233 \\
        Fine-tuning only on Affordance dataset & 0.5965 & 0.3805 & 0.0744 \\
        Fine-tuning only on InterAfford & 0.7049 & 0.3914 & 0.1134 \\
        \rowcolor{highlightrow} 
        \textbf{Two-stage Fine-tuning (Ours)} & \textbf{0.8223} & \textbf{0.5170} & \textbf{0.2036} \\
        \bottomrule
    \end{tabular}
\end{table}

\begin{table}[t]
\centering
\small
\renewcommand{\arraystretch}{1.1} 
\caption{Ablation on the reconstruction performance of Hierarchical Decoupled Quantization. ``w/o decoupled" jointly quantizes human and object motions. ``w/o hierarchical" removes the Root/Body/Hand decomposition. ``w/o geometry" excludes geometric features in object-motion encoding.}
\label{tab:ablation_quantization}

\begin{tabularx}{\linewidth}{X c c c c}
\toprule
Methods & FID$\downarrow$ & MPJPE$\downarrow$ & C\%$\uparrow$ & Coll\%$\downarrow$ \\
\midrule

w/o decoupled & 
4.625 & 0.054  & 0.348 & 0.196\\

w/o hierarchical & 
0.591 & 0.044  & 0.396 & 0.216\\

w/o geometry & 
0.370 & 0.025  & 0.452 & 0.200\\

\rowcolor{highlightrow} 
\textbf{Our Tokenizer} & 
\textbf{0.213} & \textbf{0.024}  & \textbf{0.454} & \textbf{0.186}\\

\bottomrule
\end{tabularx}
\end{table}

\begin{figure}[t]
    \centering
    \includegraphics[width=0.95\linewidth]{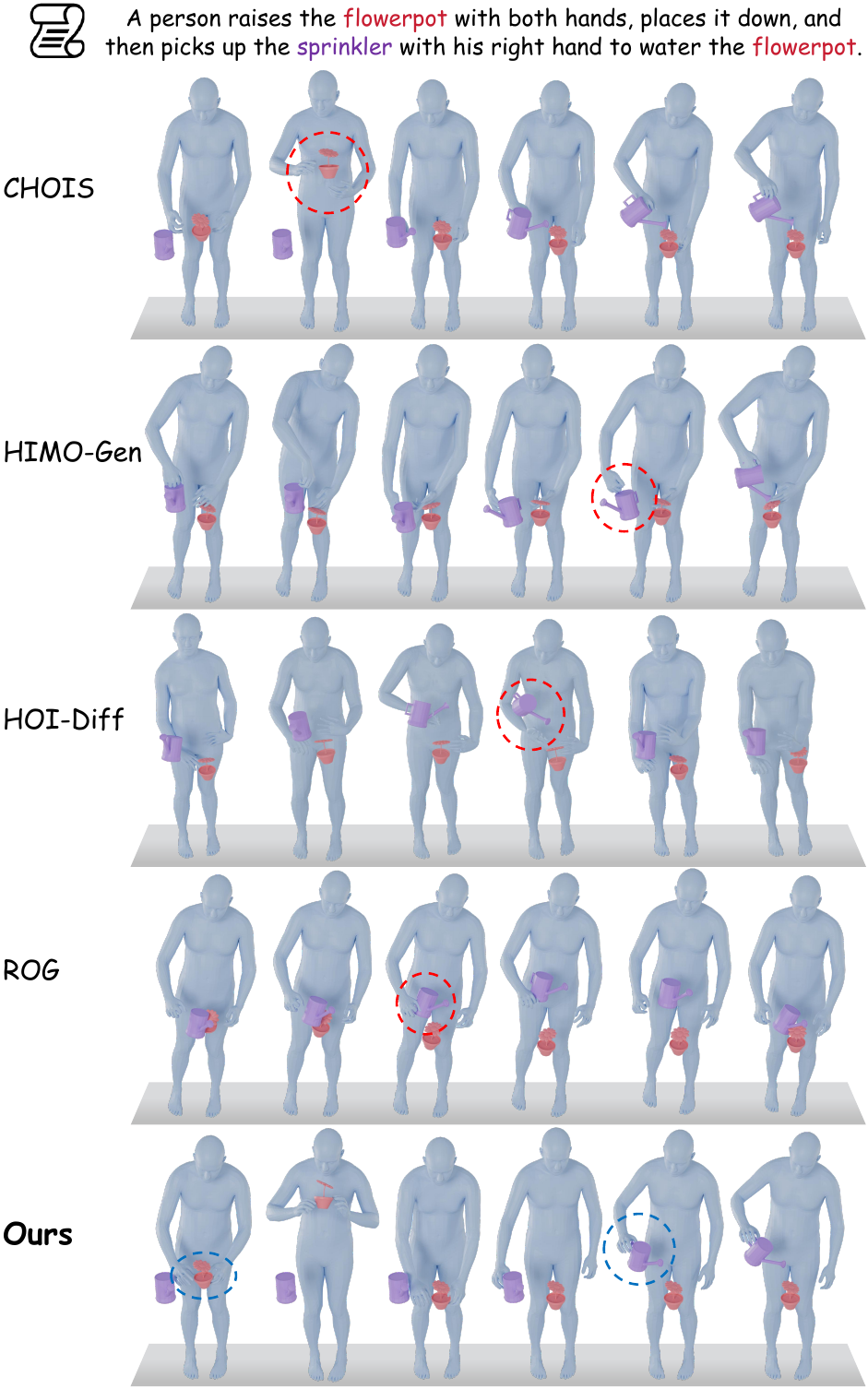}
    \caption{Qualitative comparisons with SOTA methods on HIMO dataset.}
    \label{fig:vis_1}
\end{figure}

\begin{figure*}
    \centering
    \includegraphics[width=0.88\textwidth]{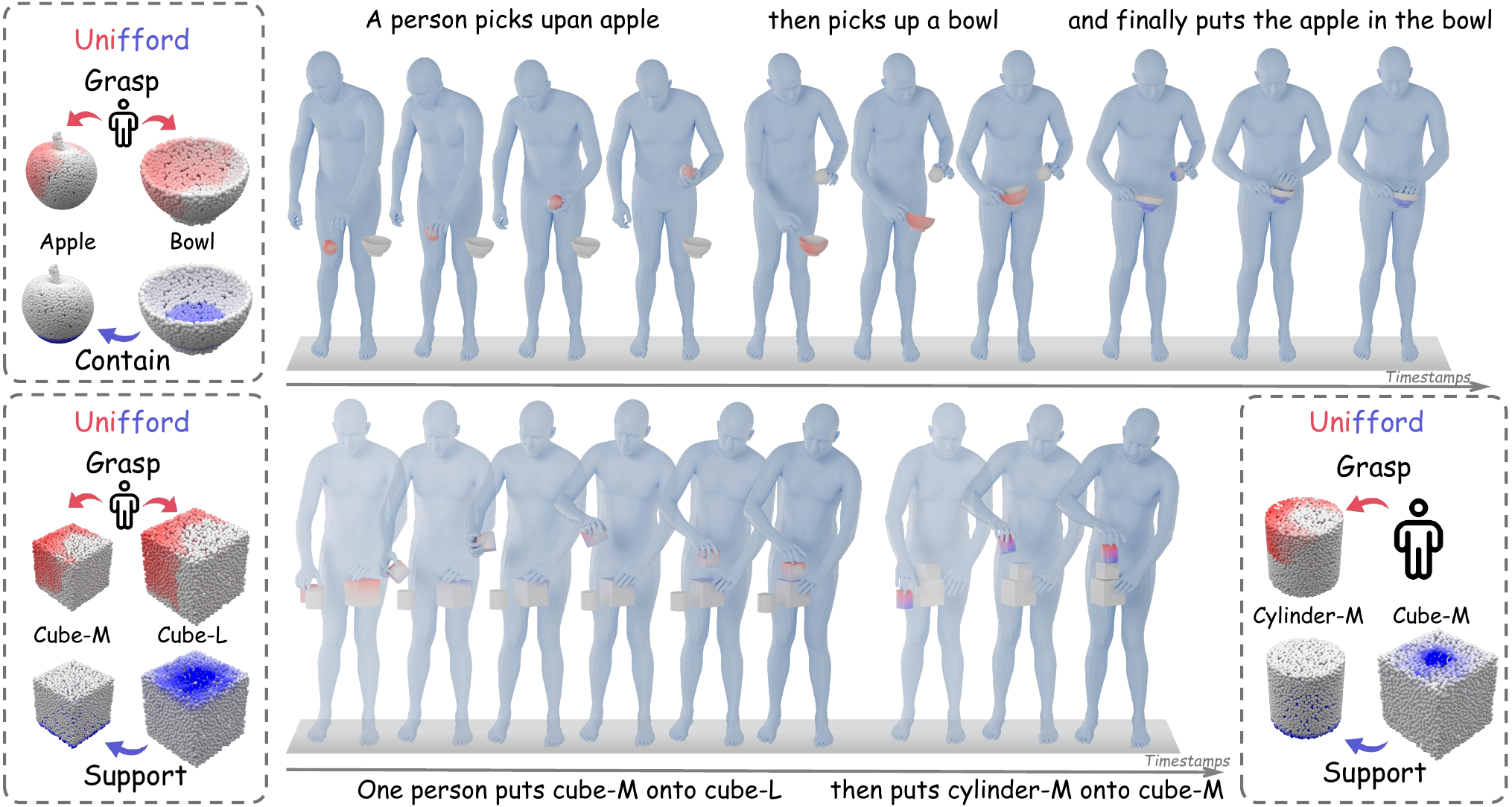}
    \caption{Visualization of the predicted affordances and the resulting human and multi-object interactive motions. The predicted Unifford maps are illustrated on the left or right, and the corresponding generated sequences are listed in the same line, demonstrating robust performance in handling sequential interactions with multi-objects and multi-actions.}

    \label{fig:vis_2}
    
\end{figure*}

\textbf{Ablation on MLLM Fine-tuning Strategies.} We first validate the necessity of the Two-stage Curriculum Training strategy within the Unified Affordance Reasoner (UAR). We conducted ablation studies comparing different training configurations, evaluating performance via Area Under the Curve (AUC), Similarity (SIM), and Intersection Over Union (IOU) metrics as shown in Table~\ref{tab:ablation_mllm}. The results indicate that the pre-trained ShapeLLM backbone yields the lowest performance (AUC: 0.4813), confirming a lack of necessary domain knowledge. Furthermore, single-stage fine-tuning on either the Affordance dataset or InterAfford proves insufficient, resulting in suboptimal IOU scores ($\le$ 0.1134) due to the inability to simultaneously capture geometric grounding and relational semantics. In contrast, our Two-stage Fine-tuning strategy significantly outperforms other configurations (AUC: 0.8223, IOU: 0.2036), demonstrating that transferring knowledge from general object grounding to specific interaction reasoning is crucial for robust unified affordance prediction.

\textbf{Ablation on Hierarchical Decoupled Quantization.}
We then demonstrate the advantages of the proposed hierarchical decoupled strategy. Specifically, we examine coarse-grained holistic representations, implemented by either jointly encoding human–object motions or removing the body-part hierarchy. As shown in Table~\ref{tab:ablation_quantization}, using such representations leads to a sharp drop in reconstruction fidelity (FID rises to 4.625). These results suggest that holistic encoding is suboptimal for HOI, as global motions(e.g., root trajectories) often overwhelm fine-grained hand manipulation. In contrast, our hierarchical decoupled design preserves reconstruction fidelity across multiple scales. Additionally, the increased collision percentage after removing geometry encoding confirms that explicitly coupling rigid motion with physical topology is crucial for safe interactions.

\textbf{Ablation on Hierarchical Autoregressive Modeling.} We further evaluate the effectiveness of our generative framework in Table~\ref{tab:ablation_autoregressive}. Replacing the coarse-to-fine generation backbone with a flat next-token prediction baseline leads to a substantial performance drop, highlighting the importance of modeling the global-to-local dependency for coherent coordination.
\textcolor{black}{To disentangle the contribution of within-frame hierarchical autoregression from multi-pass sequential generation, we evaluate a cascaded variant: it first generates the complete root-and-body sequence, then generates the hand sequence conditioned on it, and finally generates the object motions conditioned on the generated human motion. It yields a substantially higher FID and a lower Top-3 R-Precision, suggesting that sequentially generating complete motion levels is less effective at capturing fine-grained frame-level coordination than our within-frame hierarchical autoregression.}
Additionally, due to the removal of the entity embeddings, the method displays worse perceptual \wrt human parts and objects, resulting in a notable degradation in both motion and interaction quality. 

\textbf{Ablation on Unified Affordance.} As shown in Table~\ref{tab:ablation_autoregressive}, the ablation of the OCIA module leads to a significant degradation in R-Precision. This phenomenon confirms that the unified affordance from UAR provides essential semantic and geometric guidance for text–interaction alignment.
Specifically, removing the object-object affordances increases the contact rate but simultaneously causes a marked increase in collisions. 
This reveals that the model blindly pursues contact goals by sacrificing \textcolor{black}{geometric consistency}, without Unifford to coordinate the scene.

\begin{figure*}[t]
    \centering
    \includegraphics[width=0.94\textwidth]{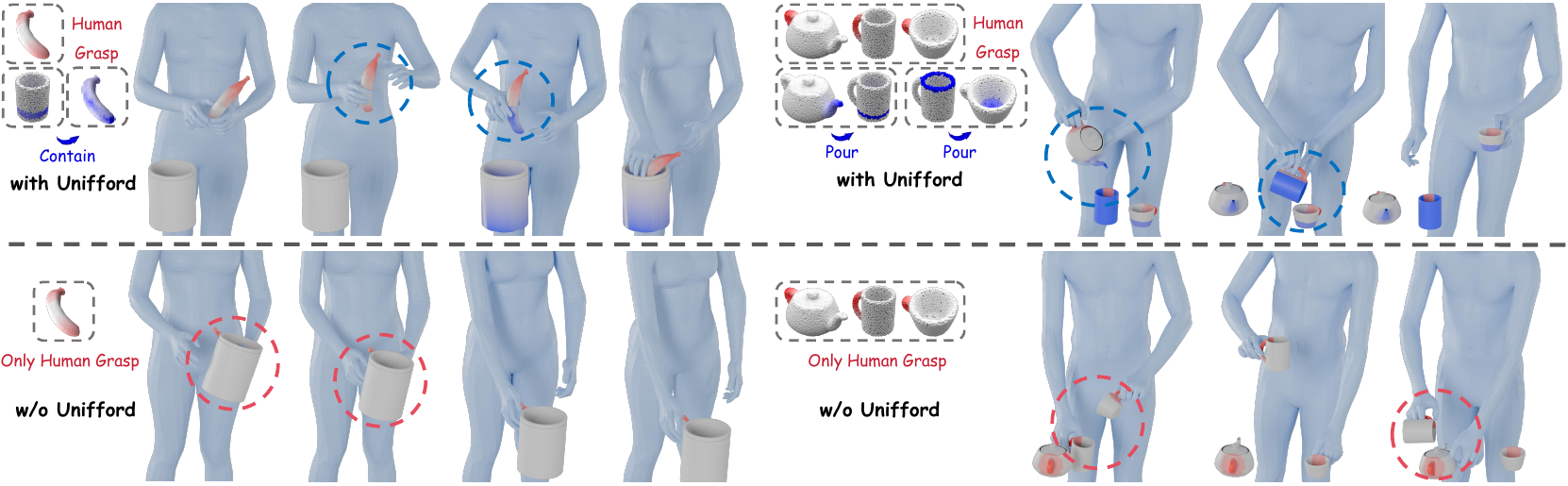}
    \vspace{-0.7em}
    \caption{Visual ablation on "picking, peeling, and placing a banana" (left) and "holding a teapot, pouring tea into a mug, transferring it to a teacup, then drinking from the teacup" (right). Top: correct execution. Bottom: failures, with the trashcan sticking to the hand and pouring applied to the wrong object.}
    \label{fig:ablation}
\end{figure*}

\begin{figure}
    \centering
    \includegraphics[width=0.95\linewidth]{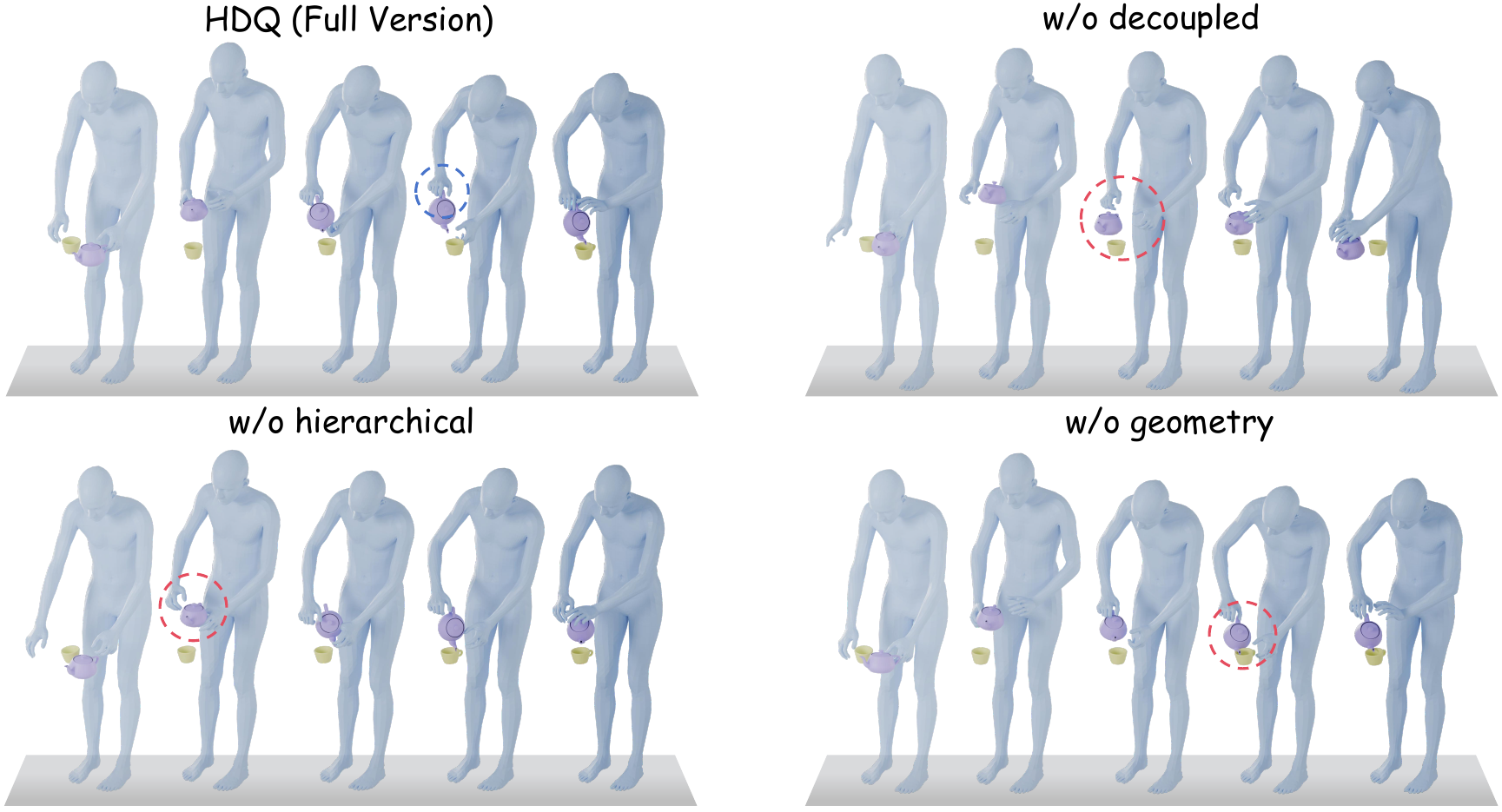}
    \caption{Visualized results of the ablation on Hierarchical Decoupled Quantization}
    \vspace{-0.7em}
    \label{fig:ablation_hdq}
\end{figure}

\begin{figure}
    \centering
    \includegraphics[width=0.95\linewidth]{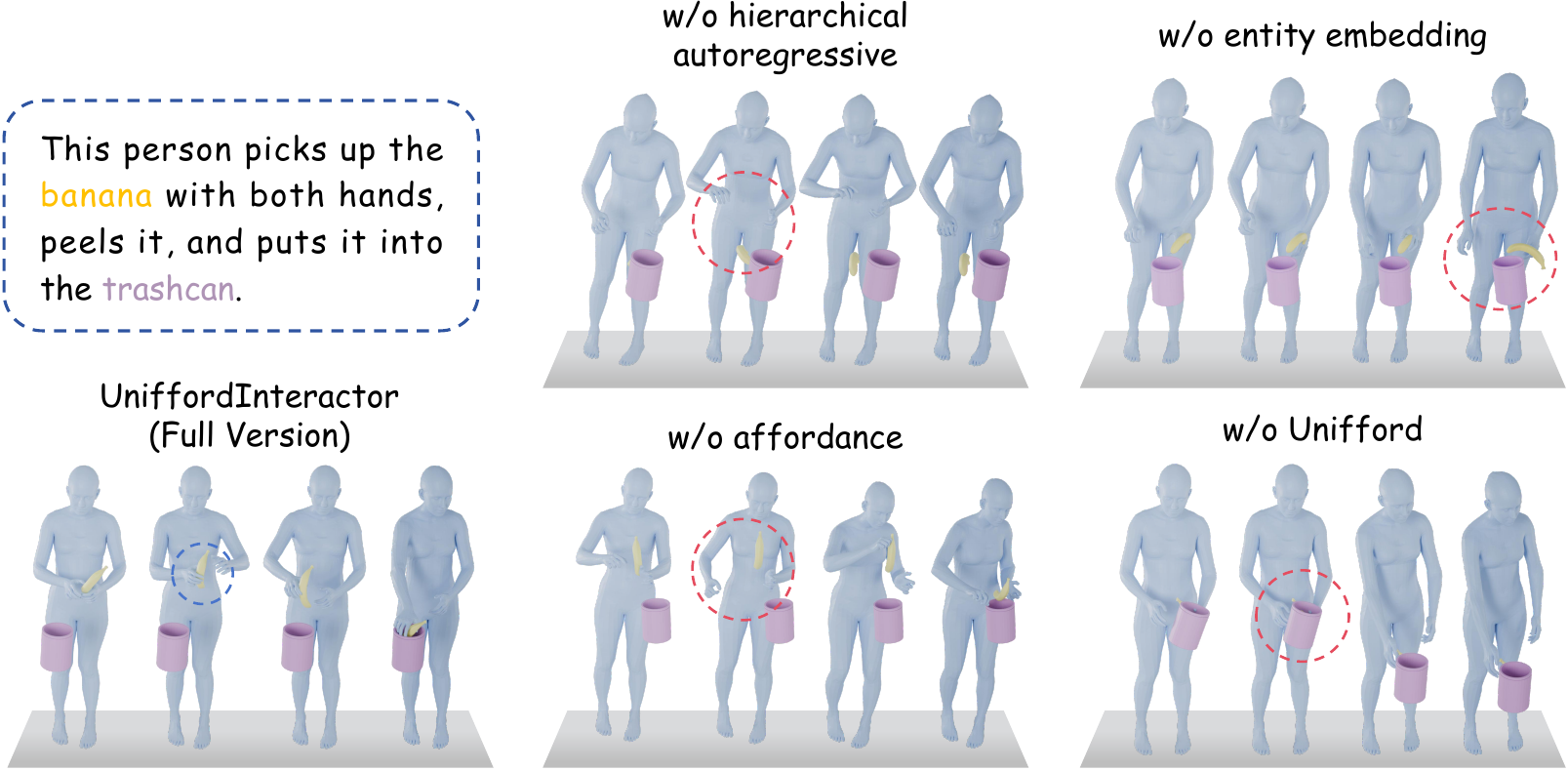}
    \caption{Visualized results of the ablation on Hierarchical Autoregressive Modeling}
    \label{fig:ablation_ham}
\end{figure}

\subsection{Qualitative Evaluation}

\textbf{Comparison with SOTA methods.} 
Figure~\ref{fig:vis_1} demonstrates our superiority in multi-object scenarios, where baselines often struggle with long-horizon planning, leading to incomplete actions or geometric artifacts (e.g., floating grasps). 
In contrast, our UniffordInteractor generates complete and \textcolor{black}{kinematically coherent} sequences. Empowered by Unifford, our method ensures precise hand–object contact and consistent spatial arrangement.
Figure~\ref{fig:vis_2} visualizes additional Unifford predictions within multi-step interaction sequences. By adding global reasoning, our model captures cascading dependencies among multiple entities. In the stacking task (bottom row), Unifford correctly maps Cube-M’s bottom surface to Cube-L and its top surface to Cylinder-M.

\textbf{Qualitative Ablations.} We provide visual comparisons to demonstrate the necessity of our core designs, including the unified reasoning mechanism and the generative components.

We first demonstrate the efficacy of unified affordance in Figure~\ref{fig:ablation}. Guided by Unifford (top row), the model jointly leverages “Grasp” and “Contain” affordances to coordinate interactions among the human, banana, and trashcan. It also correctly aligns object–object relations in more complex sequences, such as pouring from a teapot into a mug and then transferring it to a teacup. In contrast, without object–object affordances (bottom row), the model fails to capture spatial relationships, leading to errors such as the trashcan sticking to the hand or applying the pouring action to an incorrect object.

Furthermore, we visualize the impact of components within our Hierarchical Interaction Generator in Figure~\ref{fig:ablation_hdq} and Figure~\ref{fig:ablation_ham}. For quantization, our full version Hierarchical Decoupled Quantization achieves high-fidelity reconstruction, whereas ablated variants exhibit obvious geometric artifacts such as object floating and severe interpenetration (Figure~\ref{fig:ablation_hdq}). Similarly, removing the hierarchical autoregressive modeling disrupts the structural coordination, resulting in inconsistent motions and misaligned contacts (Figure~\ref{fig:ablation_ham}). This demonstrates that our full HIG framework is essential for yielding interactions with \textcolor{black}{improved contact consistency and fewer geometric artifacts}.



\begin{figure}
    \centering
    \includegraphics[width=0.99\linewidth]{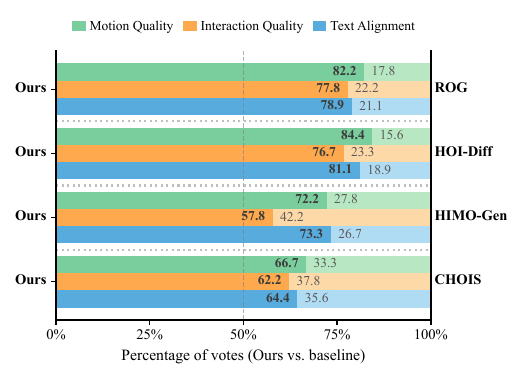}
    \caption{User study results showing the percentage of votes preferring UniffordInteractor over each baseline method under side-by-side comparisons. Results are reported separately for motion quality, interaction quality, and text alignment.}
    \label{fig:user_study_result}
\end{figure}

\textbf{User Study.} We conducted a user study to further evaluate the perceptual quality of synthesized interactions. A total of 18 participants with diverse academic backgrounds were recruited for the study. Participants were presented with side-by-side comparisons of motions generated by different methods under the same text prompt. For each comparison, two motion sequences were displayed simultaneously. 

We selected 30 text prompts, and for each prompt, all five methods generated one motion sequence. Each trial compared the result of our method against one baseline under the same prompt. Each participant evaluated 20 randomly sampled trials, with balanced coverage across the four baselines, resulting in 360 comparison trials in total and approximately 90 comparisons per baseline.

Specifically, participants evaluated the motion pairs along three dimensions: \textbf{(1) Motion Quality}, assessing the naturalness and smoothness of the motion itself; \textbf{(2) Interaction Quality}, focusing on the \textcolor{black}{geometric consistency} of human–object interactions, including contact consistency and penetration artifacts; and \textbf{(3) Text Alignment}, measuring how well the motion matches the semantic content of the text description.
The order of motion pairs and the left/right positions of different methods were randomized to avoid presentation bias. Participants were required to make a forced choice for each evaluation dimension. As shown in Figure~\ref{fig:user_study_result}, participants consistently preferred UniffordInteractor over all baseline methods across motion quality, interaction quality, and text alignment. The preference trends indicate that our method produces more natural and smoother motions, more \textcolor{black}{kinematically coherent} human–object interactions, and motions that better align with the input text descriptions, demonstrating overall improvements in perceptual quality.

\section{Conclusion}
In this paper, we propose UniffordInteractor, a novel framework designed to generate realistic long-horizon human and multi-object interaction sequences. To overcome the limitations of isolated interaction modeling, we design the Unified Affordance to explicitly unify Human-Object and Object-Object relations. To this end, we establish a Reasoning-then-Interacting paradigm in which the Unified Affordance Reasoner infers global spatial priors to guide the Hierarchical Interaction Generator. In this fashion, the generated interactions are semantically consistent and \textcolor{black}{kinematically coherent}. Extensive experiments demonstrate that our framework outperforms SOTA counterparts with superior fidelity and geometric consistency.
In the future, we will pay more attention to long-term human and multi-object interaction, and complex human-scene interaction.

\bibliographystyle{ACM-Reference-Format}
\bibliography{sample-bibliography}

\end{document}